\documentclass[final,5p,times,twocolumn]{elsarticle}

\usepackage{graphicx} 

\usepackage{subcaption}
\usepackage{amssymb}
\usepackage{amsmath}

\usepackage{float}

\usepackage{hyperref}
\usepackage{cleveref}
\usepackage{comment}
\usepackage{booktabs}  

\usepackage{multirow}
\usepackage{tabularx} 

\usepackage[nolist,nohyperlinks,printonlyused]{acronym}

\journal{Ad Hoc Networks}



\begin{document}


\begin{acronym}

\acro{ACK}[ACK]{Acknowledgment}
\acro{AI}[AI]{Artificial Intelligence}
\acro{AP}[AP]{Access Point}
\acro{API}[API]{Application Programming Interface}
\acro{AODV}[AODV]{Ad Hoc On Demand Distance Vector}
\acro{B.A.T.M.A.N.}[B.A.T.M.A.N.]{Better Approach To Mobile Ad-hoc Networking}
\acro{BSA}[BSA]{Basic Service Area}
\acro{BSS}[BSS]{Basic Service Set}
\acro{BSSID}[BSSID]{Basic Service Set Identifier}
\acro{CCK}[CCK]{Complementary Code Keying}
\acro{CLI}[CLI]{Command Line Interface}
\acro{CNN}[CNN]{Convolutional Neural Network}
\acro{CRR}[CRR]{Child Registration Request}
\acro{CSMA/CA}[CSMA/CA]{Carrier Sense Multiple Access with Collision Avoidance}
\acro{DDS}[DDS]{Data Distribution Service}
\acro{DDSNN}[DDSNN]{Decentralized Distributed Sequential Neural Network}
\acro{DHCP}[DHCP]{Dynamic Host Configuration Protocol}
\acro{DM}[DM]{Data Message}
\acro{DNN}[DNN]{Deep Neural Network}
\acro{DODAG}[DODAG]{Destination Oriented Directed Acyclic Graph}
\acro{DSDV}[DSDV]{Destination-Sequenced Distance-Vector}
\acro{DSSS}[DSSS]{Direct-Sequence Spread Spectrum}
\acro{DTSN}[DTSN]{Distributed TinyML Sensor Network}
\acro{ESS}[ESS]{Extended Service Set}
\acro{ETX}[ETX]{Expected Transmission Count}
\acro{FRU}[FRU]{Full Routing Update}
\acro{GPSR}[GPSR]{Greedy Perimeter Stateless Routing}
\acro{HAL}[HAL]{Hardware Abstraction Layer}
\acro{HEED}[HEED]{Hybrid Energy-Efficient Distributed Clustering}
\acro{HERMES}[HERMES]{Heterogeneous Application-Enabled Routing Middleware for Edge-IoT Systems}
\acro{IBSS}[IBSS]{Independent Basic Service Set}
\acro{IEEE}[IEEE]{Institute of Electrical and Electronics Engineers}
\acro{IoT}[IoT]{Internet of Things}
\acro{IP}[IP]{Internet Protocol}
\acro{LEACH}[LEACH]{Low-Energy Adaptive Clustering Hierarchy}
\acro{LLN}[LLN]{Low Power and Lossy Networks}
\acro{LSTM}[LSTM]{Long Short-Term Memory}
\acro{MAC}[MAC]{Medium Access Control}
\acro{MANET}[MANET]{Mobile Ad Hoc Network}
\acro{MIMO}[MIMO]{Multiple Input Multiple Output}
\acro{MLP}[MLP]{Multilayer Perceptron}
\acro{MonM}[MonM]{Monitoring Message}
\acro{MQTT}[MQTT]{Message Queuing Telemetry Transport}
\acro{MPR}[MPR]{Multi-Point Relay}
\acro{MU-MIMO}[MU-MIMO]{Multi-User Multiple Input Multiple Output}
\acro{MwM}[MwM]{Middleware Message}
\acro{NN}[NN]{Neural Network}
\acro{OFDM}[OFDM]{Orthogonal Frequency-Division Multiplexing}
\acro{OFDMA}[OFDMA]{Orthogonal Frequency-Division Multiple Access}
\acro{OSI}[OSI]{Open Systems Interconnection}
\acro{PDR}[PDR]{Parent Discovery Request}
\acro{PHY}[PHY]{Physical Layer}
\acro{PIO}[PIO]{PlatformIO}
\acro{PIR}[PIR]{Parent Info Response}
\acro{PRN}[PRN]{Parent Reset Notification}
\acro{PRU}[PRU]{Partial Routing Update}
\acro{OLSR}[OLSR]{Optimized Link State Routing}
\acro{QAM}[QAM]{Quadrature Amplitude Modulation}
\acro{QoS}[QoS]{Quality of Service}
\acro{RPL}[RPL]{Routing Protocol for Low-Power and Lossy Networks}
\acro{RNN}[RNN]{Recurrent Neural Network}
\acro{RSSI}[RSSI]{Received Signal Strength Indicator}
\acro{RTT}[RTT]{Round-Trip Time}
\acro{SDK}[SDK]{Software Development Kit}
\acro{SOM}[SOM]{Self-Organizing Map}
\acro{SSID}[SSID]{Service Set Identifier}
\acro{STA}[STA]{Station}
\acro{TBA}[TBA]{Topology Break Alert}
\acro{TRN}[TRN]{Topology Restored Notice}
\acro{TCP}[TCP]{Transmission Control Protocol}
\acro{TWT}[TWT]{Target Wake Time}
\acro{UDP}[UDP]{User Datagram Protocol}
\acro{WLAN}[WLAN]{Wireless Local Area Network}
\acro{WPA2}[WPA2]{Wi-Fi Protected Access 2}
\acro{WSAN}[WSAN]{Wireless Sensor Actuator Network}
\acro{WSN}[WSN]{Wireless Sensor Network}

\end{acronym} 

\begin{frontmatter}



\title{HERMES: Heterogeneous Application-Enabled Routing Middleware for Edge-IoT Systems\tnoteref{ack}}

\tnotetext[ack]{This work is funded by national funds through FCT — Fundação para a Ciência e a Tecnologia, I.P., under projects/supports UID/6486/2025 (\href{https://doi.org/10.54499/UID/06486/2025}{https://doi.org/10.54499/UID/06486/2025}) and UID/PRR/6486/2025 (\href{https://doi.org/10.54499/UID/PRR/06486/2025}{https://doi.org/10.54499/UID/PRR/06486/2025}).}

\author{Jéssica Consciência\corref{cor}} 
\ead{jessicatavares@tecnico.ulisboa.pt}
\cortext[cor]{Corresponding author.}

\author{António Grilo}
\ead{antonio.grilo@inov.pt}

\affiliation{organization={INESC INOV, Instituto Superior Técnico, Universidade de Lisboa},
             city={Lisbon},
             postcode={1000-029},
             country={Portugal}}

%

%


\begin{abstract}

The growth of the Internet of Things has enabled a new generation of applications, pushing computation and intelligence toward the network edge.
This trend, however, exposes challenges, as the heterogeneity of devices and the complex requirements of applications are often misaligned with the assumptions of traditional routing protocols, which lack the flexibility to accommodate application-layer metrics and policies.
This work addresses this gap by proposing a software framework that enhances routing flexibility by dynamically incorporating application-aware decisions.
The core of the work establishes a multi-hop Wi-Fi network of heterogeneous devices, specifically ESP8266, ESP32, and Raspberry Pi 3B.
The routing layer follows a proactive approach, while the network is fault-tolerant, maintaining operation despite both node loss and message loss.
On top of this, a middleware layer introduces three strategies for influencing routing behavior: two adapt the path a message traverses until arriving at the destination, while the third allows applications to shape the network topology.
This layer offers a flexible interface for diverse applications.
The framework was validated on a physical testbed through edge intelligence use cases, including distributing neural network inference computations across multiple devices and offloading the entire workload to the most capable node.
Distributed inference is useful in scenarios requiring low latency, energy efficiency, privacy, and autonomy.
Experimental results indicated that device heterogeneity significantly impacts network performance.
Throughput and inference duration analysis showed the influence of the strategies on application behaviour, revealed that topology critically affects decentralized performance, and demonstrated the suitability of the framework for complex tasks.

\end{abstract}


\end{frontmatter}



\section{Introduction}
\label{sec:intro}

The number of \ac{IoT} devices has grown exponentially since first surpassing the global population in 2008~\cite{history_iot},
reaching approximately 18.8 thousand million devices today with projections estimating 41.1 thousand million devices by 2030~\cite{iot_stats}.

This expansion brings new opportunities but also significant challenges.
As deployment costs decrease and device diversity rises, \ac{IoT} networks have become increasingly heterogeneous and complex.
This diversity extends beyond hardware capabilities to the variety of applications that depend on them, each with distinct requirements and objectives.

A key challenge for such applications arises from the limitations of the traditional cloud centric model, where raw sensor data is transmitted to remote servers for processing.
As networks scale, this approach becomes increasingly inefficient due to high communication overhead, energy constraints, privacy concerns, and latency.
These limitations have motivated a shift towards in network processing, where locality of data is leveraged to improve overall system performance.

Consequently, simply forwarding packets along the lowest cost path may not align with higher-level goals.
Many applications require the ability to route messages through nodes with greater processing capacity so that raw data can be converted into meaningful information before transmission.
Furthermore, this requirement exposes a structural gap: some of the information necessary for routing decisions does not reside within the network layer but at higher levels.
Factors such as available processing capacity or node reliability can play a decisive role in improving performance, yet they remain inaccessible to traditional routing protocols.

To address these limitations, this work proposes \acs{HERMES} (\acl{HERMES})\acused{HERMES}, a multi-hop routing library that seamlessly integrates heterogeneous devices and dynamically adapts to incorporate higher-level requirements.
By bridging the gap between network and application layers, \ac{HERMES} enhances routing flexibility, providing a tailored networking approach for complex \ac{IoT} applications.
The library is validated through edge computing scenarios, demonstrating two use cases: distributing \ac{NN} inference computations across multiple devices and offloading the entire workload to the most capable node.

Distributed \ac{NN} inference shows strong potential across various domains with distinct operational needs.
In energy- and latency-sensitive scenarios, local processing enables real-time responsiveness while reducing communication overhead~\cite{edge-ai-survey}, supporting predictive maintenance~\cite{DDSNN} and infrastructure monitoring.
It also facilitates rapid on-site decision-making in wildfire detection and disaster response.
In privacy-sensitive sectors such as healthcare~\cite{distributed_inference_healthcare} and finance, keeping data local allows on-premise models to operate securely.
Finally, in remote environments~\cite{remote_satellite_iot} like offshore platforms or mines, distributed inference ensures autonomous operation by enabling edge nodes to detect and respond to critical events.

The main contributions of this work relative to the state of the art are the following:
\begin{enumerate}
  \renewcommand{\theenumi}{(\arabic{enumi})}
  \item The development of a multi hop routing layer tailored for real microcontrollers and designed to ensure reliable operation in dynamic environments, supported by a \ac{HAL} that facilitates the seamless integration of new device types.

  \item A middleware layer that provides three distinct strategies, allowing applications to define custom metrics and policies to tailor the network behaviour to their specific objectives, publicly available at~\cite{HERMES}.

  \item The design of an architecture for distributed \ac{NN} inference deployed across heterogeneous devices in accordance with their computational capabilities.
\end{enumerate}

The remainder of this document is structured as follows.
\Cref{sec:backg} reviews the state of the art.
\Cref{sec:imple} presents the architecture of the system.
\Cref{sec:resul} details the experimental results.
Finally, \Cref{sec:concl} presents the conclusions and outlines directions for future work.


\section{State Of The Art}\label{sec:backg}
The state of the art relevant to this work spans four main areas: routing protocols for Edge-IoT devices, Wi-Fi-based multi-hop communication, middleware solutions for \acp{WSN}, and approaches that distribute \ac{NN} computations across multiple devices.

\subsection{Routing for Edge-IoT Devices}\label{sec:routing}
%
Routing is a fundamental process in network communication, responsible for selecting and maintaining optimal paths for data transmission between a source and a destination.
The best path is the one that satisfies all supplied constraints and has the lowest cost concerning specific metrics~\cite{RFC6551_metrics}.

Before the emergence of wireless networks, routing protocols in wired environments were traditionally categorized as either distance-vector or link-state~\cite{2-Routing_in_adhoc_survey}.
Distance-vector protocols compute routes to all destinations using information obtained only from one-hop neighbors.
In contrast, link-state protocols maintain a global view of the network topology by exchanging detailed link information among all nodes.
Although these traditional algorithms do not scale well in large ad hoc networks~\cite{2-Routing_in_adhoc_survey}, many routing protocols build upon their foundational principles.
Consequently, some works~\cite{1-Routing_in_adhoc_survey} classify ad hoc routing protocols according to the type of state information they maintain: topology based protocols, derived from link-state concepts, such as \ac{OLSR}~\cite{OLSR}; and destination based (or next-hop) protocols, derived from distance-vector principles, such as \ac{DSDV}~\cite{DSDV} and \ac{RPL}~\cite{RFC6550}.

The most common and widely accepted taxonomy distinguishes between proactive, reactive, and hybrid protocols.
Proactive (or table-driven), protocols maintain up-to-date routing tables at all times, ensuring immediate route availability at the cost of higher control overhead due to continuous updates.
Reactive (or on-demand) protocols, like \ac{AODV}~\cite{AODV} and \ac{B.A.T.M.A.N.}~\cite{BATMAN}, establish routes only when necessary, minimizing overhead but introducing initial latency during route discovery.
Hybrid protocols combine both approaches to balance responsiveness and efficiency.

Routing protocols can also be categorized based on network organization.
Flat routing assigns equal responsibility to all nodes.
Hierarchical routing introduces differentiated roles, such as cluster heads, to enhance scalability and energy efficiency.
Examples include \ac{LEACH}~\cite{LEACH} and \ac{HEED}~\cite{HEED}, which organize the network into clusters to reduce energy consumption and communication overhead.

Another emerging concept is geographical routing, where packets are forwarded based on the geographic position of the destination rather than using logical addresses or maintaining end to end paths.
Protocols such as \ac{GPSR}~\cite{GPSR} make forwarding decisions in a hop-by-hop manner using positional information, thereby eliminating the need for global topology knowledge and minimizing communication and memory overhead~\cite{Geographical_survey}.

Additionally, some authors~\cite{4-Routing_in_adhoc_survey} propose further categories, including multipath routing, which establishes multiple redundant paths between nodes, multicast and geographical multicast routing, designed to support group communication at the routing layer; and power-aware routing, which incorporates energy consumption metrics into path selection to extend network lifetime.

\subsection{Wi-Fi based Multi-Hop Networks}\label{subsec:wifi-multi-hop}


Wi-Fi typically operates in a star topology, where one or more \acp{STA} connect to a central \ac{AP}.
Extending this model to multi-hop communication, where nodes forward data among themselves, presents challenges for embedded devices, such as the ESP32 and ESP8266, mainly due to their limited memory, processing power, and energy resources.
Although Wi-Fi ad hoc mode would intuitively support such peer-to-peer communication, it remains largely unsupported on these chipsets because of security concerns and the preference of most \ac{IoT} applications for the simplicity and compatibility of infrastructure mode.
Consequently, existing works have explored alternative mechanisms to enable multi-hop communication over Wi-Fi on such constrained platforms.

Among the existing solutions, Painless Mesh~\cite{painless_mesh} and ESP-WIFI-MESH~\cite{espressif_wifi_mesh} represent more comprehensive approaches, both establishing tree topologies where each node operates simultaneously as an \ac{AP} and a \ac{STA}.
In contrast, simpler networking models have also been proposed, such as~\cite{wimp}, which employs a Directed Diffusion-inspired routing protocol with a Raspberry Pi sink node, and~\cite{esp8266_mqtt_mesh}, which adopts an MQTT-based approach where ESP8266 devices communicate through a central broker.

All these works share a common foundation: Wi-Fi multihop functionality is enabled through the softAP feature, which allows devices to act as \acp{AP}.
Each device therefore assumes two roles: as an \ac{AP}, it accepts connections from other nodes (i.e., their children), and as an \ac{STA}, it connects to a parent \ac{AP}.
In this way, each device creates its own \ac{WLAN} while connecting to the \ac{WLAN} of its parent, effectively forming a multihop topology composed of overlapping individual \acp{WLAN}.
Each node manages its own \ac{WLAN}, which can be configured with an \ac{SSID}, password, and \ac{WPA2} security.
This dual operation effectively emulates ad hoc behavior while still using infrastructure mode.

From an implementation perspective, most works~\cite{painless_mesh,wimp,esp8266_mqtt_mesh} use the Arduino Wi-Fi libraries~\cite{arduino_wifi_ESP32,arduino_wifi_ESP8266} to manage softAP functionality, relying on \ac{TCP}~\cite{painless_mesh,esp8266_mqtt_mesh} or \ac{UDP}~\cite{wimp} for inter-node communication.
These libraries inherently require the use of \ac{TCP} or \ac{UDP} at each link, introducing additional protocol overhead.
In contrast, ESP-WIFI-MESH~\cite{espressif_wifi_mesh} the native Wi-Fi stack from Espressif, enabling direct packet encapsulation within Wi-Fi data frames and thereby eliminating \ac{TCP}/\ac{UDP} dependencies.
However, this approach increases development complexity and reduces portability, as ESP32 and ESP8266 devices require different SDKs (ESP-IDF~\cite{espidf} and ESP8266 RTOS SDK~\cite{esp8266_RTOS}, respectively).
Consequently, Arduino-based solutions remain more widely adopted.

\subsection{WSN Middleware}\label{subsec:wsn-middleware}

Middleware acts as an intermediary software layer positioned between the operating system and the application layer, abstracting hardware and network complexities while providing essential services such as data aggregation, synchronization, and communication management.
Since WSNs are typically constrained by limited energy, dynamic topologies, and hardware heterogeneity, this layer must remain lightweight and adaptive to ensure dependable operation and allow developers to focus on application logic rather than low-level details.

Several architectural paradigms have influenced middleware design~\cite{middleware_survey,middleware_survey_healthcare}.
Event-based systems rely on asynchronous publish/subscribe communication, service-oriented architectures expose devices as discoverable services to promote modularity, and virtual machine-based solutions provide portability and safety through a uniform execution interface.
Agent-based middleware focuses on autonomous agents capable of migration and adaptation, while tuple-space and database-oriented models enable data-centric coordination through shared or queryable abstractions.
Some middleware systems are further specialized for specific domains to meet application-level \ac{QoS} requirements.

Among early data-centric middleware solutions, TinyDSM~\cite{tinyDSM} is a distributed shared-memory system designed to enhance data reliability and availability in WSNs.
It establishes a cooperative data storage mechanism that enables sensor nodes to replicate their data within a configurable, locality-bound neighborhood.
PD-MidI~\cite{pattern_based_middleware} extends event-driven design into a layered architecture that supports asynchronous data collection and processing, while embedding built-in privacy, security, and anonymization features.

Recent research has incorporated semantic technologies to address heterogeneity and enable intelligent, context-aware processing.
The work~\cite{sim_middleware} integrates semantic annotation, blockchain, and \ac{AI} based analysis within a unified framework to ensure data integrity and enable intelligent processing, primarily in healthcare applications.
Similarly,~\cite{sedia} applies a semantic approach to urban environments, harmonizing heterogeneous \ac{IoT} data sources into a unified and context-enriched structure.

Other middleware solutions address specific operational challenges.
Reliable TinyDDS~\cite{RTDDS} extends the \ac{DDS} publish/subscribe model with a three-tier \ac{QoS} mechanism that introduces Partial Reliability to guarantee the delivery of critical data while conserving network resources.
Conversely, the work in~\cite{middleware_smart_campus} targets large-scale smart campus deployments and employs a modular microservices architecture with dynamic protocol translation and adaptive resource management, ensuring scalability and interoperability in dense, heterogeneous environments.

In summary, existing middleware approaches emphasise modularity, adaptability, and lightweight operation, which align with the goals of this work.
Unlike cloud-based solutions~\cite{sim_middleware,sedia,middleware_smart_campus}, the middleware proposed in this paper is embedded at the network level, similar to~\cite{tinyDSM,RTDDS}, operating close to the distributed mesh rather than remotely above it.
Another difference lies in scope.
Whereas many solutions target specific concerns such as reliability~\cite{RTDDS}, data availability~\cite{tinyDSM}, or semantic annotation~\cite{sim_middleware,sedia}, the middleware developed in this work provides mechanisms that allow applications to influence lower layers and adapt their behaviour to the application context.

\subsection{Distributed Neural Network Computation}\label{subsec: distributed-neural-networks-computation-in-wsns}

The convergence of \ac{AI} and \ac{IoT} motivates distributed intelligence, where machine learning models are partitioned and processed collaboratively across edge devices.
Neural Networks and distributed computing share a natural analogy: each device acts as a neuron or group of neurons performing local computation, while communication links represent inter-neuron connections.
Approaches range from deploying complete models on individual nodes~\cite{ann_per_node_forest}, distributing Neural Networks across more capable edge devices~\cite{DNN_edge_raspberry,distributted-inference-cnn,priority_aware_MDI,early_exit_MDI}, to frameworks optimized for resource-constrained devices~\cite{AI_WSN,WSN_ANN,DTSN,DDSNN}, as well as studies focusing on neuron-to-device mapping~\cite{WSN_as_a_computer_framework_NN} and computational complexity of such systems~\cite{Complexity_mlp_wsn}.

Early work~\cite{ann_per_node_forest} proposes a detection system for forest fires, where each TelosB sensor node runs a complete \ac{MLP}.
In contrast, other approaches focus on offloading and partitioning models across more powerful edge devices.
For instance, ~\cite{DNN_edge_raspberry} introduces CoopAI, a cooperative edge computing system where \ac{IoT} devices send data to a gateway that partitions a DNN across Raspberry Pi edge devices to address network partitioning for parallel computation.
Other works~\cite{distributted-inference-cnn, priority_aware_MDI, early_exit_MDI} explore optimized inference on high-performance edge devices like NVIDIA Jetsons.
These employ techniques such as partitioning based on device resources, priority-aware task allocation, and adaptive early-exit strategies to optimize distributed inference.

For resource constrained devices, several approaches distribute Neural Network computation directly across the \ac{WSN}.
AI-WSN~\cite{AI_WSN} employs a \ac{SOM} \ac{NN} where each node represents a single neuron, evaluated through network simulations.
Similarly, WSN-ANN~\cite{WSN_ANN} proposes an architecture with a supervisory node coordinating network operations while other nodes function as individual neurons in a Hopfield Neural Network, applied to solve graph problems.
The authors of ~\cite{WSN_as_a_computer_framework_NN} propose a framework to dynamically map hidden neurons onto an \ac{IoT} network, formulating the optimal placement as an integer linear program to minimize communication costs.
Moving beyond simulation, more recent efforts demonstrate practical implementations on physical hardware: DTSN~\cite{DTSN} enables distributed training and inference by partitioning \acp{NN} into layer blocks across ESP32 devices communicating via Bluetooth mesh, while DDSNN~\cite{DDSNN} employs a pipelined approach with layer-wise partitioning of pre-trained models across multiple microcontrollers over Wi-Fi.

Extending the discussion toward theoretical evaluation, the authors of ~\cite{Complexity_mlp_wsn} provide a thorough complexity analysis of training a \ac{MLP} within a WSN where each node acts as a single neuron.
Their study concludes that while space and time complexity scale manageably, message complexity emerges as the fundamental limitation for scalability.

In summary, the explored approaches reveal a clear trajectory from centralized computation towards increasingly decentralized and collaborative paradigms that fully exploit network devices.
In~\cite{ann_per_node_forest}, each node independently executes its own \ac{NN}, resulting in redundant computation instead of leveraging distribution to share the workload and conserve energy.
In contrast, \cite{DNN_edge_raspberry,distributted-inference-cnn,priority_aware_MDI,early_exit_MDI} explore distributed inference but target powerful edge devices, with memory capacities in the gigabyte range.
This stands in sharp contrast to the devices considered in this work, where the least capable node is limited to just 64 KB of SRAM.
Approaches like~\cite{AI_WSN,WSN_ANN,WSN_as_a_computer_framework_NN} distribute computation across the \ac{WSN}, though they remain limited to software simulations without real-world deployment.
More practical evaluations appear in~\cite{DTSN,DDSNN}, which test distributed execution on physical devices, yet still neglect hardware heterogeneity and employ comparatively powerful microcontrollers (e.g., ESP32-S3 with 512 KB of SRAM).
Finally,~\cite{Complexity_mlp_wsn} highlights that communication cost is the main factor limiting distributed inference.
This insight is incorporated into the present work to optimize data exchange between nodes.


\section{Heterogeneous Application-Enabled Routing Middleware for Edge-IoT Systems}\label{sec:imple}
%

This section presents the development process and architecture of \ac{HERMES}.
It begins with \Cref{subsec:system-overview}, which outlines the system and its main components.
The following subsections adopt a bottom-up approach: \Cref{subsec:core-network} details the core network and implemented protocols, \Cref{subsec:middleware-layer} introduces the middleware strategies that link routing with application logic, and \Cref{subsec:application-layer} describes the application layer and its use cases for validating the proposed strategies.
In addition, \Cref{subsec:network-monitoring-tool} describes the tool specifically created to obtain the performance measurements later presented in the \Cref{sec:resul}.

\subsection{System Overview}\label{subsec:system-overview}

This work presents a multi hop routing framework for heterogeneous \ac{IoT} devices that enables seamless integration of nodes with varying computational capabilities while supporting dynamic route optimization based on application requirements.

%
%

\begin{figure*}[htbp]
\centering
\includegraphics[width=0.9\textwidth]{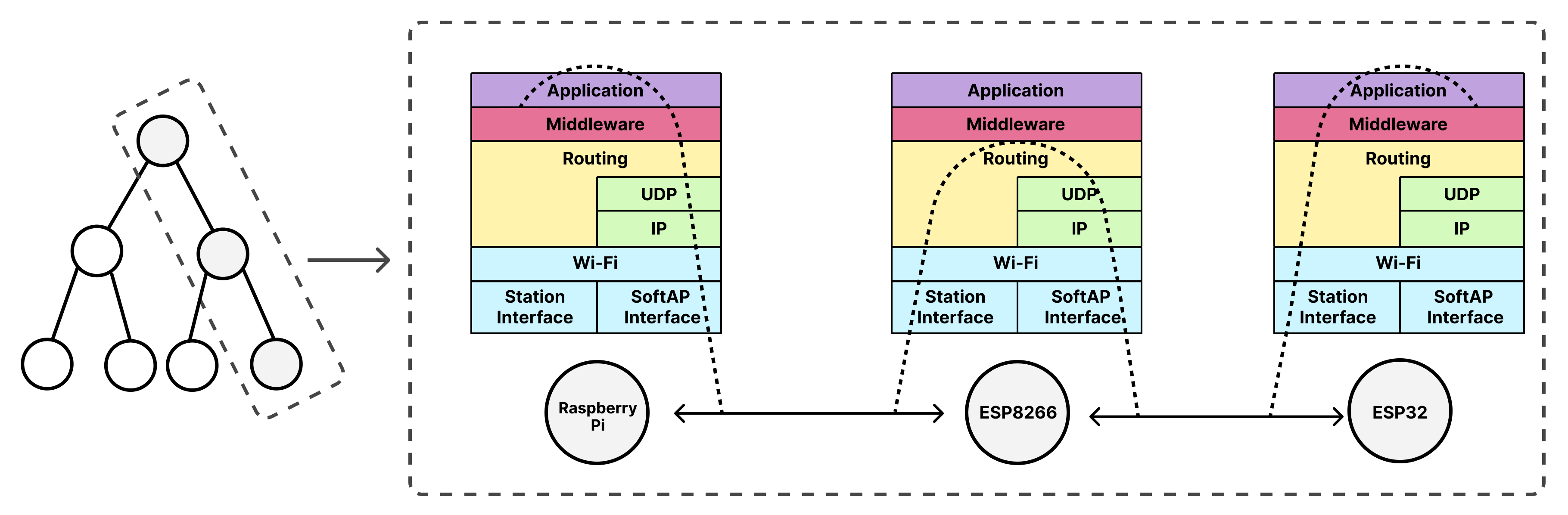}
\caption{System Architecture.}
\label{fig:system_general_architecture}
\end{figure*}

The overall system architecture is depicted in Figure~\ref{fig:system_general_architecture}.
The network adopts a tree topology, since Wi-Fi infrastructure mode only allows a connection to one parent \ac{AP}.
Each node follows the same layered stack, with the Wi-Fi layer at the base providing communication through its softAP and \ac{STA} interfaces.
Above it are the \ac{UDP} and \ac{IP} layers, responsible for enabling direct data transfer between nodes.
The routing layer sits on top of these communication layers, handling packet forwarding across the network.
Above routing, the middleware provides an abstraction interface that allows the application to influence network behavior dynamically.
Finally, the application layer contains the use-case-specific logic, leveraging middleware mechanisms to adapt system operation according to application goals.

%

The system addresses hardware heterogeneity across ESP8266, ESP32, and Raspberry Pi platforms through a \ac{HAL} that encapsulates device-specific implementations behind unified interfaces.
Each \ac{HAL} module exposes a uniform \ac{API} that remains consistent across platforms, while individual devices provide their own implementations according to their hardware capabilities.
The upper layers interact solely with these abstract interfaces and remain unaware of the underlying differences.
Platform-specific code is selected at compile time through conditional compilation, ensuring that only the correct implementation is included for each target.
This design keeps the upper layers platform-agnostic and enables straightforward horizontal scaling, since supporting new devices requires only the implementation of the corresponding HAL interface functions, without modifying upper-layer logic.

\subsection{Core Network}\label{subsec:core-network}
The core network consists of establishing and maintaining an ad hoc, multi-hop network in which nodes collaborate without centralized control.
This section presents its main components.



\subsubsection{IP/Wi-Fi Subnetwork}\label{subsubsec:4-Wi-Fi}

\begin{figure}[H]
    \centering
    \includegraphics[width=0.9\linewidth]{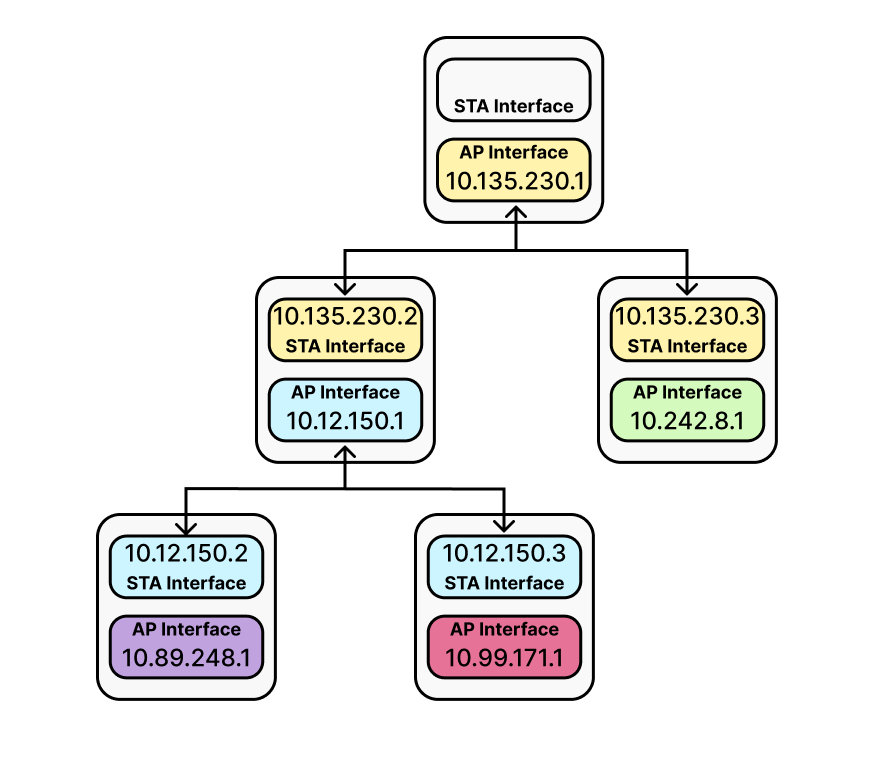}
    \caption{Resulting Network of Interconnected \acp{WLAN}, Showcasing the Dual IP Address Configuration on Each Node.}
    \label{fig:muliple_wlan_ips}
\end{figure}

Each device in the network exposes two distinct Wi-Fi interfaces: a softAP interface, which allows other devices to connect as children, and a \ac{STA} interface that connects to the \ac{AP} interface of another node, forming a parent–child relationship.
This dual-interface setup generates overlapping \ac{WLAN}, as each node creates its own \ac{WLAN} while joining another, and together these links form the complete multi-hop network, as shown in \Cref{fig:muliple_wlan_ips}.

As discussed earlier, the ESP8266 and ESP32 platforms support Wi-Fi softAP functionality through two primary approaches.
This work adopts the Arduino Wi-Fi libraries.
In contrast, for the Raspberry Pi, no equivalent library was available, so the implementation was developed using low-level system calls.

To establish its own \ac{WLAN}, each device configures its \ac{AP} interface with a specific \ac{SSID}, passphrase, static \ac{IP} address, gateway, and subnet mask.
Once the \ac{AP} is active, a \ac{DHCP} service starts automatically, assigning \ac{IP} addresses to any child node that connects.
These addresses are derived from the configured \ac{IP} and subnet mask of the \ac{AP}.

As a result, every node in the network maintains two distinct \ac{IP} addresses: one for its \ac{AP} interface and one for its \ac{STA} interface.
The \ac{AP} \ac{IP} address is statically configured during the initial setup and remains unchanged throughout the lifetime of the node.
Conversely, the \ac{STA} \ac{IP} address is dynamic and depends on the current parent to which the node is connected.
When a node switches parents, the \ac{DHCP} server of the new parent assigns it a new \ac{STA} \ac{IP} address.
This dual \ac{IP} configuration is illustrated in the \Cref{fig:muliple_wlan_ips}.

All nodes share a common \ac{SSID} prefix and password to enable discovery.
Static \ac{AP} \acp{IP} are derived from the \ac{MAC} address of the node: the first octet is fixed at 10 (private address range), the next two come from the last two bytes of the \ac{MAC}, and the final octet is 1.
The gateway \ac{IP} of each node matches its \ac{AP} \ac{IP}, and the subnet mask is 255.255.255.0 (/24).


The Wi-Fi layer defines events triggered by parent and child connections or disconnections.
These events allow the system to monitor direct links and enable upper layers to detect and respond to link breakages.

\subsubsection{Transport Protocol}\label{subsubsec:5-UDP}

\ac{UDP} was chosen for its connectionless nature, low overhead, and broadcast support.
Its lightweight design enables fast message transmission on resource-constrained devices; however, delivery guarantees and message ordering must be handled by upper layers.

\subsubsection{Routing}\label{subsubsec:routing}

The default routing protocol adopts a proactive and destination-based design, maintaining up-to-date routes to all nodes while storing only the next hop toward each destination.
This ensures that valid routes are immediately available whenever needed, avoiding the delays of on-demand discovery.
Storing only the next hop reduces memory usage, which is beneficial for devices with limited resources.
The protocol draws on concepts from \ac{DSDV}, employing sequence numbers to ensure route freshness and prevent routing loops.

Each node maintains a routing table containing an entry for every other node in the network.
Each entry specifies the \ac{AP} \ac{IP} address of the destination, the next hop, the hop distance, and a sequence number.
\ac{AP} \acp{IP} are used for consistency, as they uniquely identify nodes and remain static throughout their lifetime.
Sequence numbers play a central role in maintaining valid routes.
Nodes increment their own sequence numbers by two whenever they issue a routing update.
If a node detects a link failure to a neighbor, it marks that destination as unreachable by incrementing the sequence number of the neighbor by one (creating an odd value) and setting the hop count to infinity.
Odd sequence numbers thus serve as explicit unreachability indicators, invalidating the associated route until a subsequent even-numbered update reinstates path validity.

When a node receives routing information from a neighbor, it updates its routing table accordingly.
Routes to previously unknown nodes are added with the next hop set to the sender and the hop distance set to the advertised value plus one.
For existing routes, updates depend on the sequence number: entries with older sequence numbers are discarded, those with the same sequence number are updated only if the new path is shorter, and those with higher sequence numbers fully replace the previous entry.

Establishing and maintaining consistent routing paths across the network is achieved through periodic routing updates.
The routing layer defines two types of messages: \acp{FRU}, which contain the entire routing table, and \acp{PRU}, which carry only modified entries.

Each node periodically sends a \ac{PRU} containing only the significant changes since the last \ac{FRU}.
\acp{FRU} are sent less frequently.
If the size of a \ac{PRU} approaches that of an \ac{FRU}, a full update is sent instead.
When a node receives an update, it updates its routing table and propagates only the significant changes.
Significant changes include the addition of new nodes, link failures, or modifications to existing paths.
Minor updates, such as sequence number increments, are sent only in the next scheduled \ac{FRU}.

Because the network follows a tree topology, only a single path exists to each destination.
So if a node becomes unreachable, all routes passing through it must also be invalidated.

Finally, when sending a message to a remote destination, the node consults its routing table to determine the next hop, and the message is relayed through intermediate nodes until it reaches its target.
Each node also maintains a mapping table that translates the \ac{AP} \acp{IP} of its children into their corresponding \ac{STA} \acp{IP}, as communication with a child node always occurs through its \ac{STA} interface.

\subsubsection{Node Lifecycle}\label{subsubsec:node-lifecycle}

\begin{figure}[H]
\centering
\includegraphics[width=\linewidth]{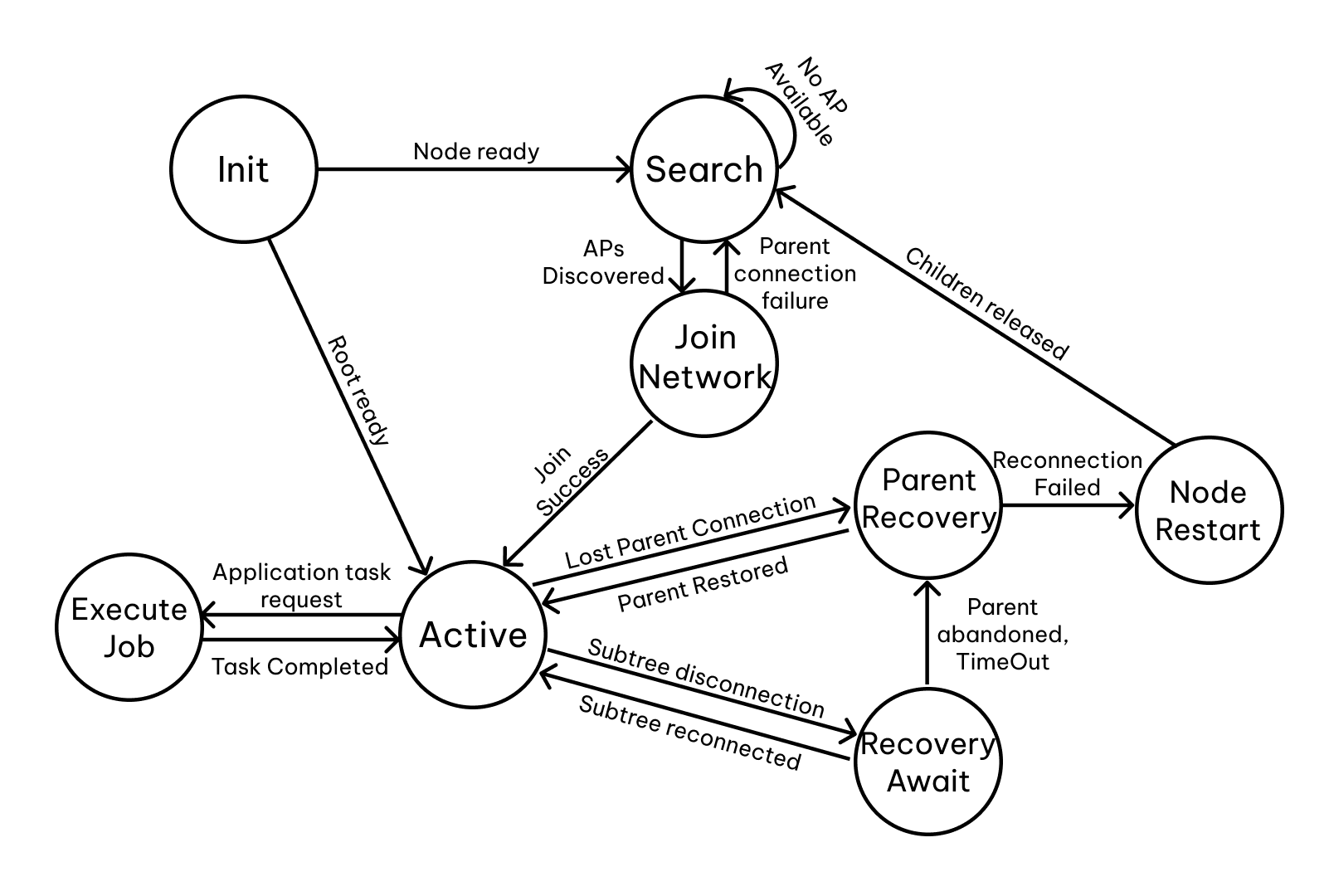}
\caption{State Machine.}
\label{fig:state_machine}
\end{figure}

The node lifecycle encompasses all operational phases, from initial network integration to adaptation following topological changes, such as parent disconnections.
It is implemented through an event-driven state machine, composed of a set of states and an event queue.
The event queue is implemented as a circular buffer, offering a fixed memory footprint, efficient event handling, and automatic overwriting of older events to prevent stalls under high load.
The state machine, illustrated in ~\Cref{fig:state_machine}, consists of eight states grouped into three functional categories: network integration (\textit{Init}, \textit{Search}, \textit{Join Network}), normal operation (\textit{Active}, \textit{Execute Job}), and recovery (\textit{Parent Recovery}, \textit{Recovery Await}, \textit{Node Restart}).

When a node enters the network, it begins in the \textit{Init} state.
Root nodes transition directly to \textit{Active}, while non-root nodes enter \textit{Search} to discover potential parents.
Once at least one parent is found, the node proceeds to \textit{Join Network}, contacts the candidate parents to gather their status, selects a preferred parent, and attempts to establish a connection.
Upon success, the node enters \textit{Active}, participating fully in network operations, including routing updates, middleware routines, and child management.
Application-level tasks are executed in \textit{Execute Job}, after which the node returns to \textit{Active}.

Recovery from losses in connections is managed through three states.
In \textit{Parent Recovery}, a node that loses its parent actively searches for a new parent while preserving its subtree.
It transitions to \textit{Active} if recovery succeeds or to \textit{Node Restart} if it fails.
In \textit{Node Restart}, the node releases all direct children, resets its internal network state, and returns to \textit{Search} to rejoin the network.

During this period, nodes within the detached subtree enter \textit{Recovery Await}, suspending normal operations but still maintaining routing mechanisms to preserve subtree integrity.
If the node that initially lost connectivity successfully reconnects to the main tree, it notifies all descendants, which then transition back to the \textit{Active} state.
If it fails to establish a new parent, its direct children exit \textit{Recovery Await} and enter \textit{Parent Recovery} to seek their own parent, while their descendants remain in \textit{Recovery Await}.
This mechanism enables detached subtrees to remain internally connected and to reintegrate gradually, layer by layer, ensuring an orderly recovery process.

The node lifecycle includes fault-tolerance mechanisms to ensure continuous operation in unreliable wireless ad hoc environments.
Incoming messages are validated to prevent system failures due to corruption, and out-of-order delivery does not compromise node or network state.
All waits for parent responses and recovery actions are time-bounded to avoid indefinite blocking.
To mitigate the loss of critical control messages, a safeguarding mechanism leverages the routing protocol; for instance, an odd sequence number for the root in a routing update can trigger the node to transition to \textit{Recovery Await}, ensuring the network state remains consistent even when alerts are lost.

\subsection{Middleware Layer}\label{subsec:middleware-layer}

The middleware acts as an intermediate layer between the application and routing layers, enabling applications to influence routing behavior.
This influence may occur either by directing packets along specific paths or by shaping the network topology to ensure that particular nodes are included in the forwarding routes.
Three distinct middleware strategies were developed to achieve these goals.

The design of the middleware layer was inspired by the strategy pattern, where each strategy implements a common interface while encapsulating its own logic to influence routing decisions.
This architecture allows the application layer to select and apply strategies without awareness of their internal mechanisms.
It also ensures that the system remains open for extension but closed for modification, as new strategies can be integrated seamlessly by implementing the defined interface.
This enables future enhancements without requiring any changes to the stable lower layers.

\subsubsection{Inject Strategy}\label{subsubsec:strategy-inject}

\begin{figure}[h]
  \centering
    \includegraphics[width=\linewidth]{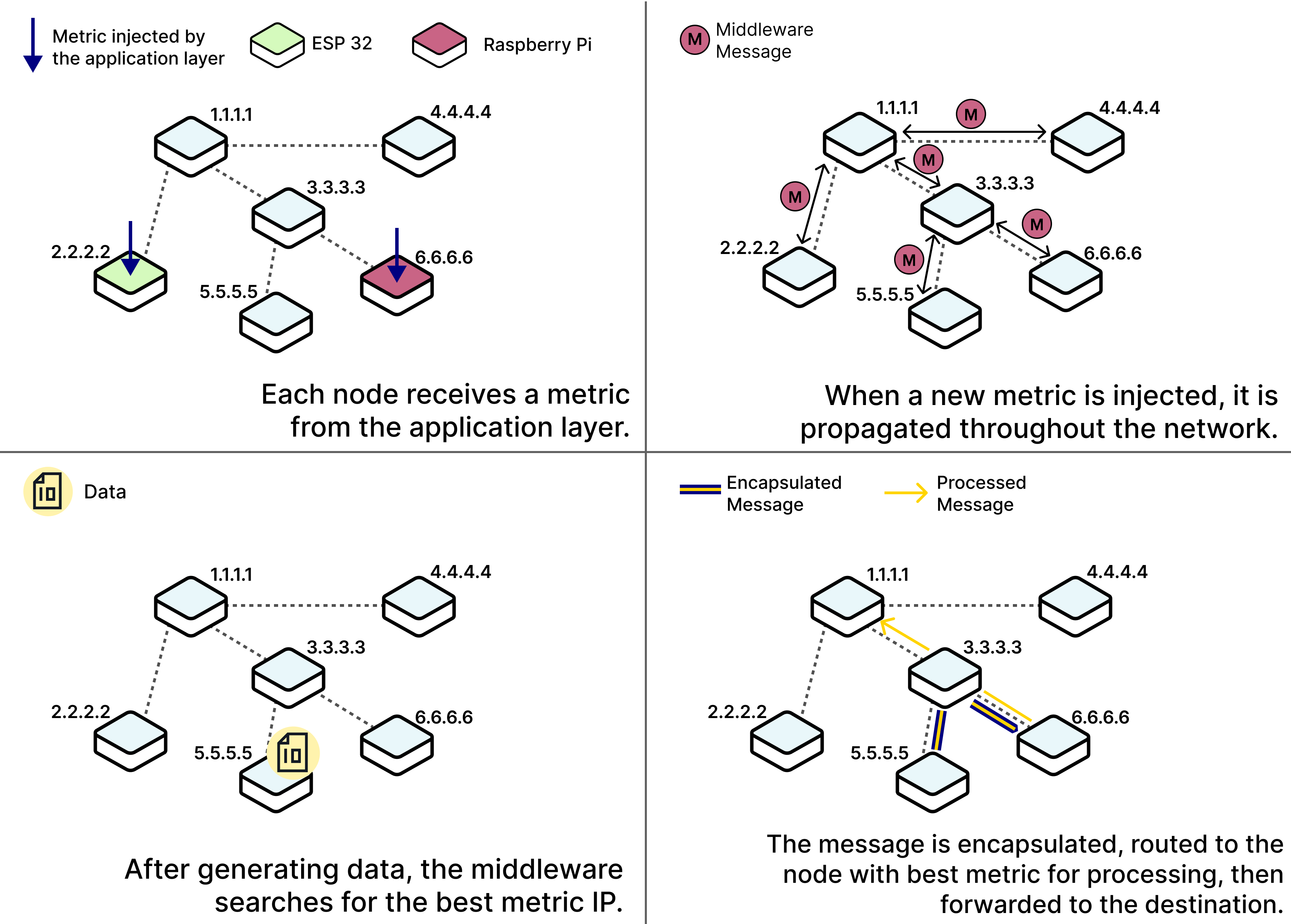}
  \caption{Inject Strategy.}
  \label{fig:strategy_inject}
\end{figure}

In the Inject Strategy, each node can be associated with a metric defined and managed by the application layer, such as processing capacity.
This metric is injected by the application into the middleware layer, which is then responsible for propagating it across the network.
By distributing these values, the application gains the ability to steer routing decisions, since the metrics provide a way to identify which nodes are most suitable for handling specific tasks.
In practice, this means the application layer can redirect data messages through an intermediate node with a more favorable metric before they continue on to the final recipient.
The strategy is illustrated in the \Cref{fig:strategy_inject}.

When a node joins the network, it reports its metric value to its parent, which then replies with the current metrics of the other nodes.
These values are periodically refreshed and adjusted to reflect application changes, ensuring an up-to-date view of node capabilities.

To initialize this strategy, the application layer provides the details of the metric structure and a comparison function for evaluating two metric values.
Metrics can therefore represent a single property or a composite of several, and the comparison logic can implement any policy the application considers relevant.
For instance, the processing capacity of a node could be combined with its current load, so that decisions consider not only maximum capacity but also the node availability.

When the application wishes to influence routing, the strategy evaluates the metrics of the available nodes according to the policy defined by the application and, if a suitable candidate is found, encapsulates the original data message inside another message addressed to the node with the optimal metric.
In this encapsulated message, the internal and final destination remains the intended recipient, while the external destination is the chosen best-metric node.
The intermediate node processes the message and then forwards the processed result to the final recipient.
Furthermore, if two nodes share equal metrics, the one with the shortest hop distance is chosen.
If no suitable node is found, the message is sent directly to its destination.

\subsubsection{Publish and Subscribe Strategy}\label{subsubsec:strategy-pubsub}
\begin{figure}[htbp]
  \centering
    \includegraphics[width=\linewidth]{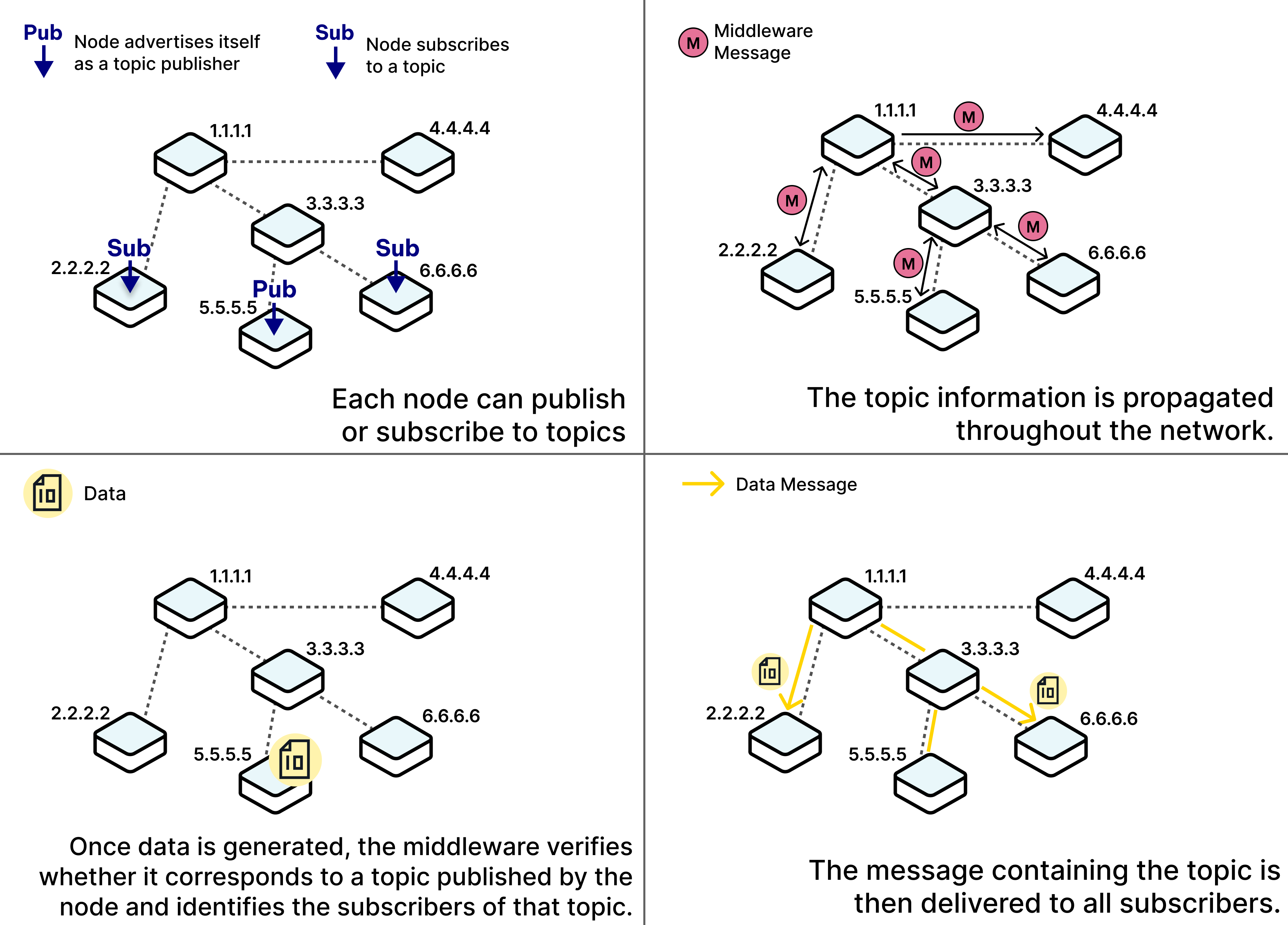}
  \caption{Publish and Subscribe Strategy.}
  \label{fig:strategy_pubsub}
\end{figure}

In the Publish and Subscribe Strategy, each node may act as a publisher or subscriber of specific topics.
A node that produces information announces itself as a publisher of a given topic, while nodes interested in receiving such information register as subscribers.
The middleware propagates information about the topics that nodes publish and subscribe to across the network.
When the application layer wishes to influence routing, the strategy checks whether a data message relates to a topic published by the node.
If so, the message is delivered to all nodes that have subscribed to that topic.
This strategy is illustrated in the \Cref{fig:strategy_pubsub}.

Within the middleware layer, topics are represented by integers, while their interpretation is entirely defined at the application layer.
This abstraction provides flexibility, as topics can correspond to any kind of information.
For example, topics may represent different data types produced by nodes, such as temperature readings, humidity values, or motion events.
Alternatively, topics may identify geographical zones in which the network is deployed, allowing certain nodes to subscribe only to information relevant to their area of interest.

When a node joins the network, it reports its published and subscribed topics to its parent, which responds with information about other nodes.
Throughout its lifecycle, a node can publish or withdraw topics and manage subscriptions.
To avoid stale information, each node periodically propagates information about its topics.

\subsubsection{Topology Strategy}\label{subsubsec:strategy-topology}

The Topology Strategy enables the application layer to influence routing by directly controlling how nodes establish their parent-child relationships.
As in the Inject Strategy, each node can be associated with a metric defined and managed by the application layer.
However, in this case, only the root node maintains a global view of all metrics and uses this information to determine how the network should be structured.

\begin{figure}[H]
  \centering
    \includegraphics[width=\linewidth]{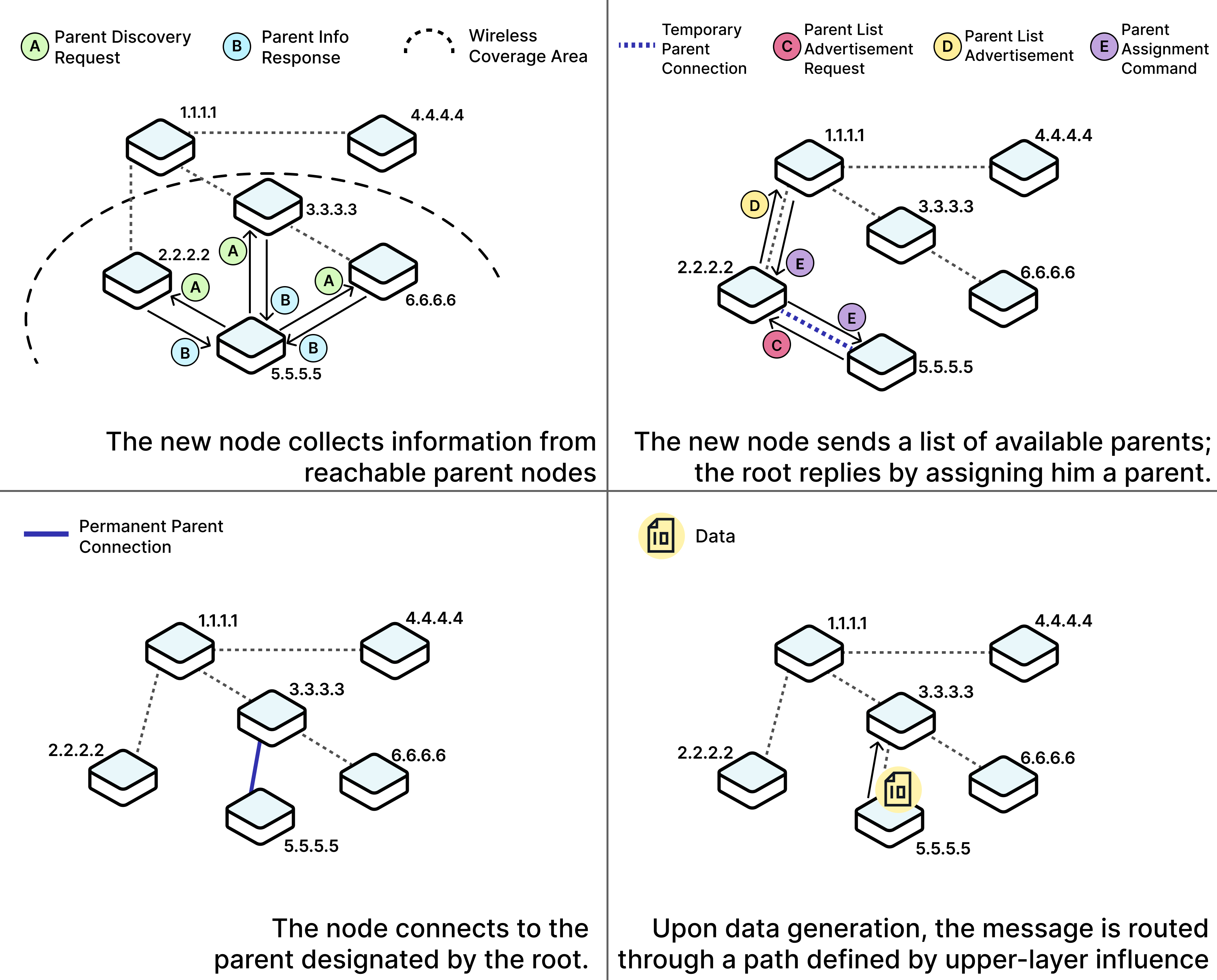}
  \caption{Topology Strategy.}
  \label{fig:strategy_topology}
\end{figure}

When a node joins the network, it first contacts all potential parents to assess their availability and status.
Then to gain provisional network access, it temporarily attaches to one of these candidates and sends a \textit{Parent List Advertisement Request} message that contains a list of all potential parents.
The temporary parent then forwards a \textit{Parent List Advertisement} message to the root, which examines the provided information and selects the definitive parent for the joining node.
The decision is communicated through a \textit{Parent Assignment Command} message, relayed back by the temporary parent, completing the integration process.
This process is illustrated in the~\Cref{fig:strategy_topology}.

Once integrated, each node periodically sends a message to the root containing its metric and the \ac{IP} address of its current parent.
This enables the root to maintain an up-to-date view of the entire topology.

The application layer initializes this strategy by providing the metric structure details and a parent selection function, which chooses the preferred parent from a list of candidates.
This function can consider both individual node metrics and the current network topology.
For example, a node that generates data requiring significant processing can be directly attached to a node with greater computational resources, thereby minimizing the number of hops between nodes that frequently exchange data and reducing overall traffic.
Alternatively, the function may isolate unstable or unpredictable nodes by assigning them to less critical parts of the network, limiting the impact of potential failures.

\subsubsection{Applicability and Comparison}\label{subsubsec:strategies-differences}

The three strategies provide ways to influence routing behaviour, but their applicability differs.
The Inject Strategy is suitable when data must pass through a node with specific characteristics, such as a security module for encryption, even if that node does not require the data for its own operations.
The Publish-Subscribe Strategy applies to structured data flows where intermediary nodes need the data, as in industrial automation systems where sensors publish measurements and controllers subscribe to initiate actions.
The Topology Strategy is most effective when network structure directly impacts performance, enabling optimizations such as connecting nodes to more trustworthy neighbors to reduce packet loss.

The key distinction between the strategies lies in the level of routing influence: Inject places decision-making at the data source, allowing it to dynamically select intermediate nodes and recipients; Publish/Subscribe shifts control to intermediary nodes, which define the data categories that pass through them; and Topology preconfigures network connectivity to inherently satisfy application requirements.

\subsection{Application Layer}\label{subsec:application-layer}
The application layer sits on top the core network and middleware components and hosts two use cases for testing the developed strategies: a distributed \ac{NN} across multiple devices and a centralized \ac{NN} executed on the device with the highest processing capacity.

\subsubsection{Scenario 1: Distributed Neural Network}\label{subsubsec:distributed_neural_network}

\begin{figure}[H]
  \centering
    \includegraphics[width=\linewidth]{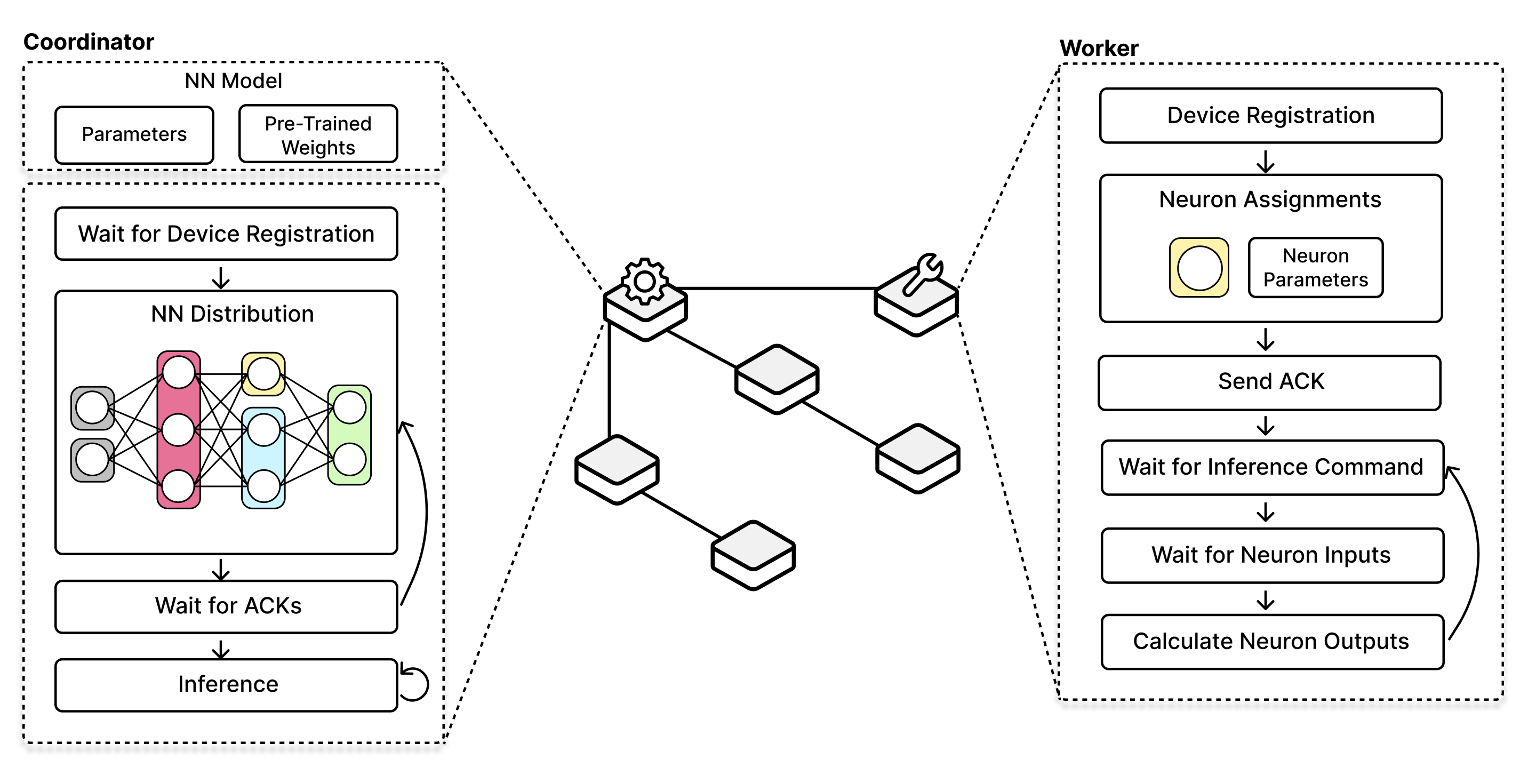}
  \caption{Architecture of the Distributed Neural Network Solution.}
  \label{fig:distributted_nn_system}
\end{figure}

The distributed \ac{NN} architecture features two node roles: a coordinator and workers.
The coordinator manages the network, distributing computations and directing data flow, while workers compute their assigned neurons.
Workers can handle neurons from any layer, and the coordinator can also act as a worker, contributing directly to computations.
Each node maintains information according to its role: the coordinator stores the complete \ac{NN} model with pre-trained weights, biases, and neuron assignments, whereas workers retain only the data needed for their assigned neurons.
The architecture of the distributed \ac{NN} is depicted in~\Cref{fig:distributted_nn_system}.

Devices register with the coordinator according to their \ac{NN} role (input generation, hidden, or output computation).
Once sufficient devices are registered, the coordinator distributes neurons based on device capabilities, assigning more neurons to more capable nodes.

Hidden-layer neurons are allocated sequentially, layer by layer, across the registered devices.
This layer-wise assignment strategy increases the likelihood that a device will compute neurons within the same layer, thereby taking advantage of shared input dependencies.
Output neurons are assigned to a single device that explicitly registers for this role; if no such device is available, the coordinator assumes responsibility for their computation.

Devices acknowledge their assignments, and the coordinator resends any unconfirmed allocations.
Once all acknowledgments have been received, the coordinator issues the command to start the inference.
Each forward pass is associated with a sequence number managed by the coordinator, the Inference Id, which ensures that devices accept and process only the outputs corresponding to the current inference cycle, thereby maintaining synchronization and data consistency across the network.
Inferences can be performed periodically.

During inference, workers compute their assigned neurons as inputs arrive.
If required inputs are missing after a specified timeout, the worker broadcasts a NACK message.
Should these inputs remain unavailable after an additional waiting period, the worker substitutes the missing values with those from the previous inference cycle, ensuring continuous operation.

\subsubsection{Scenario 2: Centralized Neural Network}\label{subsubsec:centralized_neural_network}

The centralized use case employs the same framework as the distributed \ac{NN}.
In this configuration, a single device registers as both the worker and the output node, prompting the coordinator to assign all \ac{NN} neurons to that device.

\subsubsection{Integration of Use Cases with the Middleware Strategies}\label{subsubsec:integration_of_use_cases_with_middleware}

\begin{figure}[H]
    \centering
    \includegraphics[width=\linewidth]{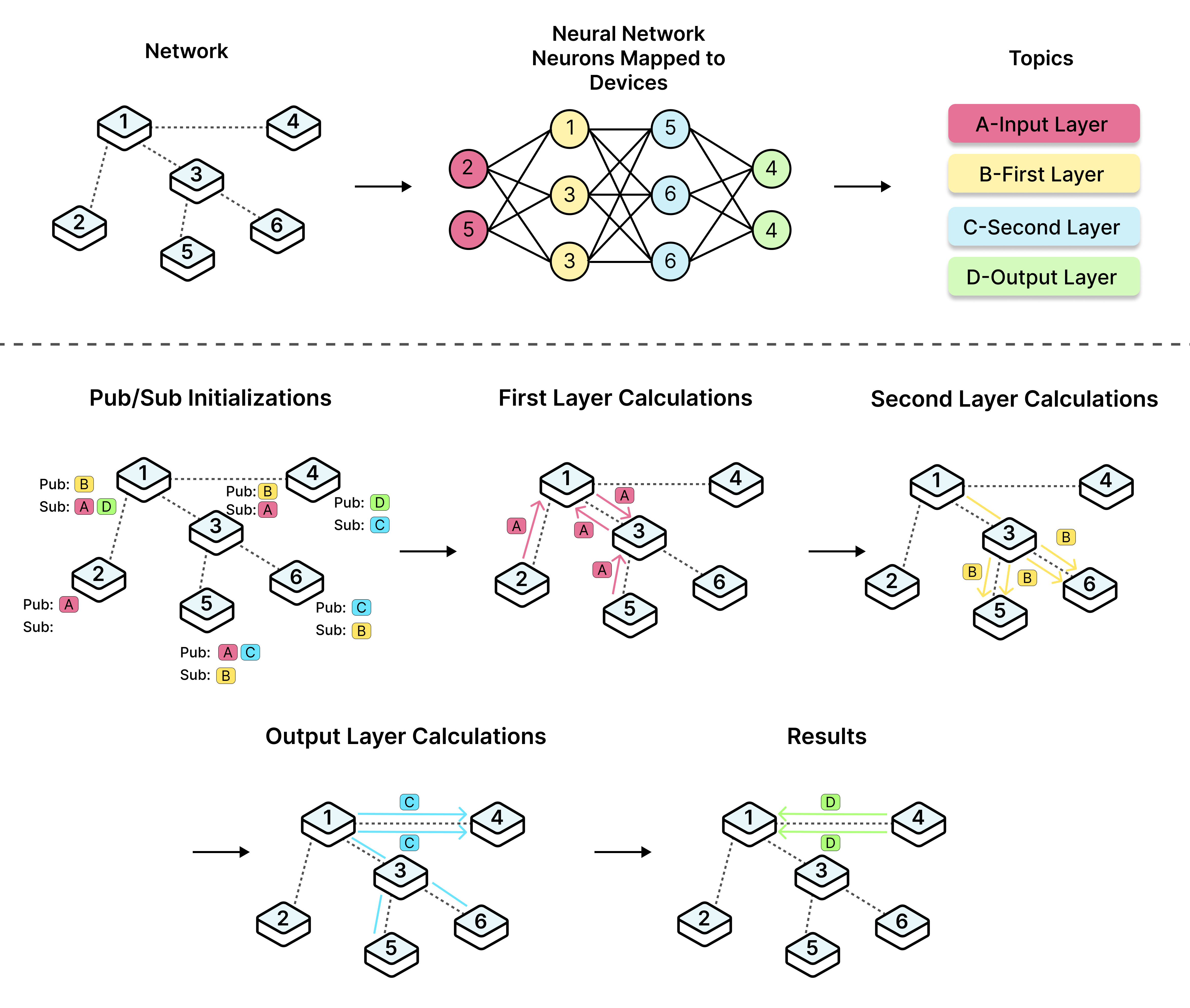}
    \caption{Integration of the Publish and Subscribe Strategy with the Distributed NN Use Case.}
    \label{fig:distributed_nn_pubsub}
\end{figure}

The distributed \ac{NN} use case was tested with both the Publish and Subscribe Strategy and the Topology Strategy.
The Publish and Subscribe Strategy proved to be a natural fit, as \acp{NN} inherently follow a structured, layered flow of information where neurons in one layer produce outputs consumed by the next.
This maps seamlessly to the publish and subscribe paradigm, with each layer represented by a distinct topic.
For instance, the input layer corresponds to topic 0, and subsequent layers use incrementing topic identifiers.
Each neuron publishes its output under the topic corresponding to its own layer and subscribes to the topics of the preceding layer.
\Cref{fig:distributed_nn_pubsub} illustrates the integration of the Publish and Subscribe Strategy with the distributed \ac{NN}, highlighting the mapping of layers to topics as well as the publications and subscriptions of the devices.

Using the Publish and Subscribe strategy, devices do not need to be explicitly informed about which others depend on their output, since the middleware layer manages all publication and subscription information.
On the application side, the only requirement is to encode a message with the neuron output value.
The middleware then extracts the topic identifier from that message and delivers it to all subscribers, abstracting all device-level addressing.

The Topology Strategy was used to explore how the network structure could be optimized to reduce the number of hops messages traverse, improving neuron-to-node communication efficiency.
Since this strategy does not assist in message forwarding, each device had to be explicitly informed of which subsequent nodes required its outputs for further computation.

In contrast, the Inject Strategy could not be effectively applied to the distributed case.
It is more suitable for non-hierarchical scenarios where data flows through intermediate processing nodes without strict layer dependencies.
Therefore, it aligns naturally with the centralized use case, where the entire Neural Network is computed on a single device.

\subsection{Network Monitoring Tool}\label{subsec:network-monitoring-tool}

To support the visualisation and evaluation of the system, the Network Monitoring Tool was developed.
This tool enables real-time observation of the network topology and facilitates the assessment of the system through customisable performance metrics.
Its operation relies on reading the serial output of the root node, which emits a special message type containing the information required by the tool.
The Network Monitoring Tool proved essential during the evaluation process, serving as the primary instrument for collecting the data presented in \Cref{sec:resul}.

%
%


\section{Experimental Results}\label{sec:resul}

This section presents the results obtained from the evaluation of the system.
The analysis is structured around two main components.
First, \Cref{subsec:core_network} evaluates the core network in isolation, without the middleware or application layers, to establish a performance baseline for the higher-level components.
Second, \Cref{subsec:middleware-strategies-evaluation} is dedicated to the evaluation of the middleware strategies, examining their effectiveness when applied to the defined use cases.

The experimental testbed was composed of five devices in total: one NodeMCU (ESP8266), three ESP32-WROVER boards, and one Raspberry Pi 3 Model B+.

\subsection{Core Network}\label{subsec:core_network}

\begin{figure}[H]
    \centering
    \includegraphics[width=\linewidth]{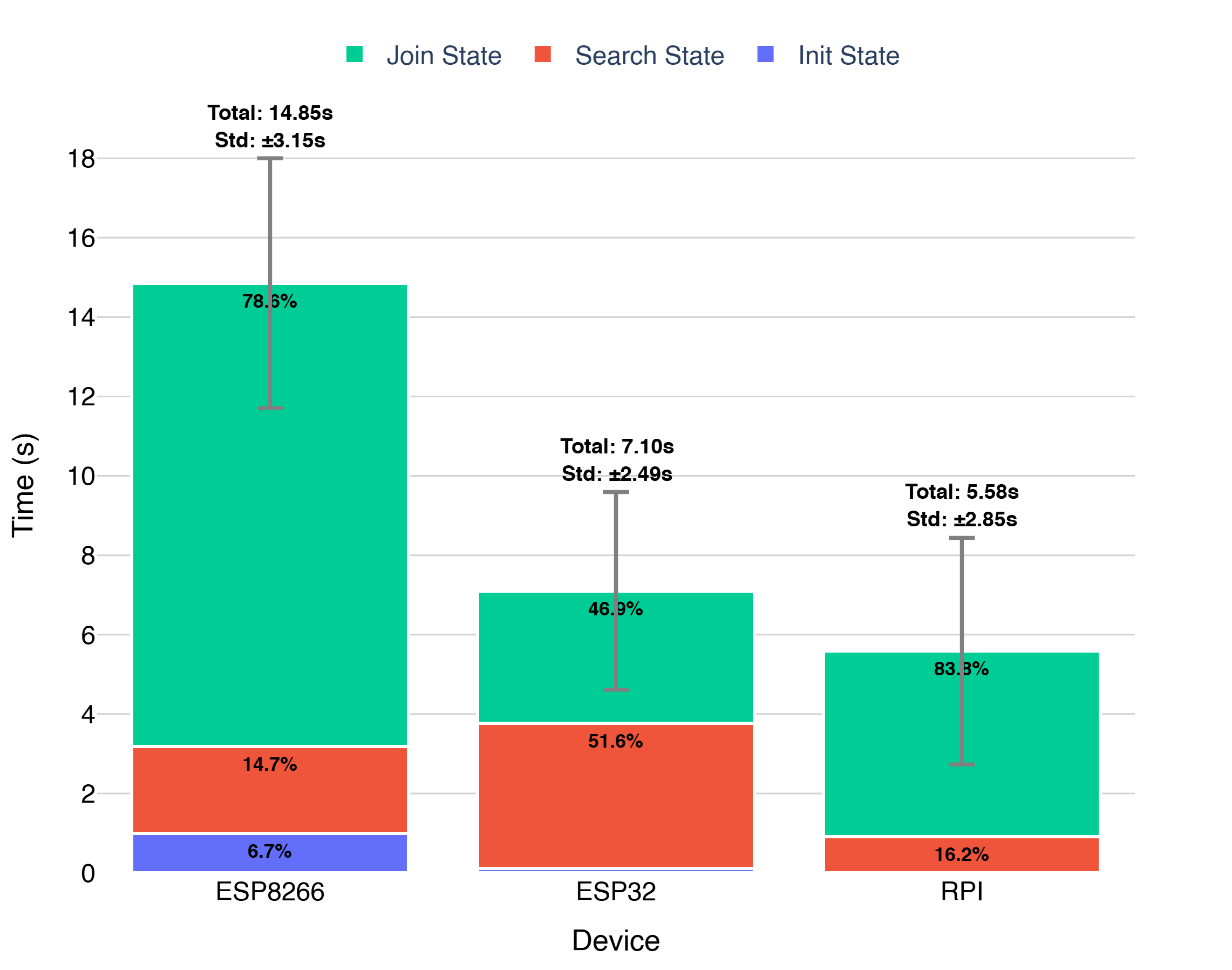}
    \caption{Network Integration Time by Device.}
    \label{fig:network-integration-time-by-device}
\end{figure}

The analysis begins with the network integration phase.
The total integration time corresponds to the period between the start of the program and the moment when the node exits the \textit{Join Network} state, encompassing the durations of the \textit{Init}, \textit{Search}, and \textit{Join Network} states.
After completing integration, each device reports its state durations to the root node.
\Cref{fig:network-integration-time-by-device} presents the average network integration times for each device type, calculated from multiple trials.
The results indicate a clear correlation between device computational power and integration speed: the ESP8266, being the least powerful device, required significantly more time to integrate, taking 109\% longer than the ESP32 and 166\% longer than the Raspberry Pi, which achieved the fastest integration.
The \textit{Join Network} state was the most time-consuming overall, although the ESP32 exhibited comparable durations in the \textit{Search} and \textit{Join Network} phases.

\begin{figure}[H]
    \centering
    \includegraphics[width=\linewidth]{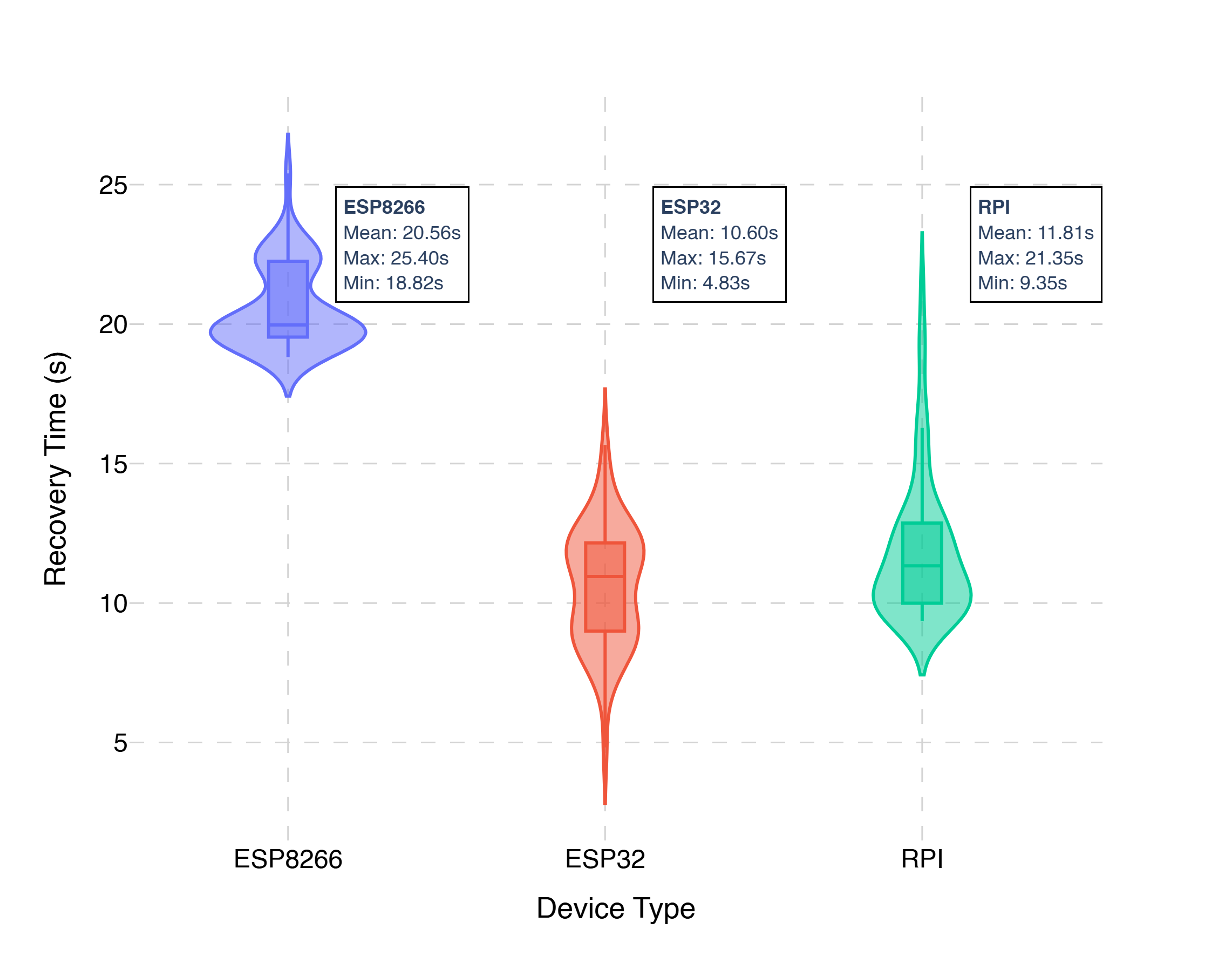}
    \caption{Parent Recovery Time by Device.}
    \label{fig:parent-recovery-time-by-device}
\end{figure}

Further measurements evaluated recovery after parent loss, illustrated in \Cref{fig:parent-recovery-time-by-device}.
Results followed the same trend, with the ESP8266 exhibiting the slowest recovery, and the ESP32 and Raspberry Pi achieving the fastest, confirming that device heterogeneity influences network responsiveness and stability.



The measured \ac{RTT} in the network increases with hop count, as expected in a multi-hop topology.
For single-hop communication, the mean latency was approximately 42.6\,ms with a standard deviation of 50.5\,ms.
At two hops, the mean latency rose to around 153.2\,ms, showing a mean incremental cost of 110.6\,ms per additional hop.
However, the variability increased significantly, with a standard deviation of 77.2\,ms.

Throughput analysis of the messages received by the root node, as shown in \Cref{tab:table-throughput-results-core-network}, indicates that routing messages dominated the traffic, reaching 1.02\,B/s with a routing update interval of 60\,s.

\begin{table}[H]
\centering
\small
\caption{Throughput Results of the Core Network.}
\begin{tabular}{lc}
\toprule
\textbf{Category} & \textbf{Value} \\
\midrule
Total Messages & 1.44\,B/s \\
Routing Messages & 1.02\,B/s \\
Lifecycle Messages & 0.20\,B/s \\
Monitoring Messages & 0.22\,B/s \\
\bottomrule
\end{tabular}\label{tab:table-throughput-results-core-network}
\end{table}

\subsection{Middleware Strategies and Application Evaluation}\label{subsec:middleware-strategies-evaluation}

At the application layer, \ac{NN} inference was performed using a 12-neuron \ac{MLP} architecture comprising 2 input neurons, 2 hidden layers of 4 neurons each, and 2 output neurons.
The ESP8266 acted as the network coordinator, while the ESP32 devices generated the input values for inference.
The \ac{NN} was not trained for any specific domain but was instead configured as a generic benchmark to validate the functionality of the application and middleware layers.

For the distributed \ac{NN} use case, the number of neurons assigned to each device matched its capacity.
Each ESP32 processed one neuron, the Raspberry Pi executed an entire hidden layer plus one additional neuron, and the coordinator node (ESP8266) handled the two output neurons.

At the middleware layer, each strategy employed a distinct configuration.
In the Topology Strategy, applied to the distributed \ac{NN} use case, each node was assigned a processing capacity metric (1 for ESP8266, 2 for ESP32, and 3 for Raspberry Pi).
The coordinator selected parent nodes by prioritising those with higher capacity.
Since nodes with greater capacity are assigned more neurons, they are more likely to produce values required by other nodes, and prioritising them as parents ensures that dependent nodes can access these values directly.
In the Inject Strategy, used in the centralized use case, only the node executing the complete \ac{NN} inference injected its metric into the middleware, signalling its role as the primary inference node.
Finally, the Publish/Subscribe Strategy, applied to the distributed \ac{NN} use case, operated with the default middleware configuration and required no additional setup.

The first aspect analyzed was the initialization phase, which is common to both the centralized and distributed \ac{NN} implementations.
The duration of this phase was defined as the time elapsed from the start of neuron assignment until the reception of all acknowledgments.
The Inject and Topology Strategies performed stably, with mean durations of 310.92\,ms and 371.20\,ms, and standard deviations of 53.35\,ms and 87.43\,ms, respectively.
The longer time for the Topology Strategy is expected due to its multi-device setup requiring each device to acknowledge its assigned neurons, whereas the Inject Strategy only required the Raspberry Pi to acknowledge all assignments.
No missing acknowledgments were observed for these strategies.

In contrast, the Publish and Subscribe Strategy showed a much higher mean initialization time of 2904.60\,ms and a standard deviation of 2689.43 ms.
This delay was caused by network congestion: upon receiving neuron assignments, nodes immediately subscribed to relevant topics and acknowledged their assignments, generating a surge of middleware control traffic that led to packet loss, including ACKs destined for the coordinator, forcing retransmissions and delaying phase completion.

\begin{figure}[H]
\begin{subfigure}[t]{\linewidth}
    \centering
    \includegraphics[width=\linewidth]{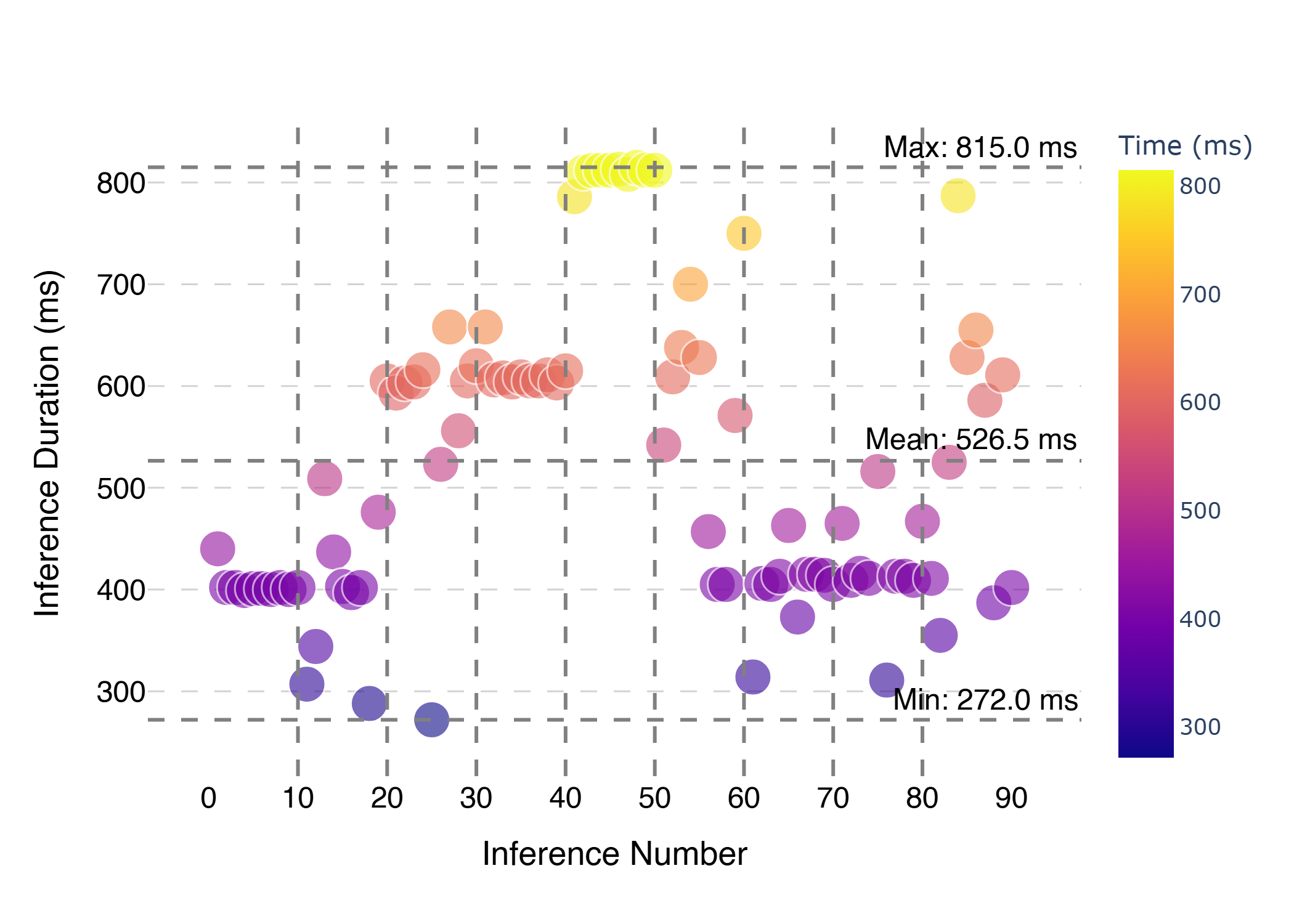}
    \caption{}
    \label{fig:pubsub-inference-duration}
\end{subfigure}
\begin{subfigure}[t]{\linewidth}
    \centering
    \includegraphics[width=\linewidth]{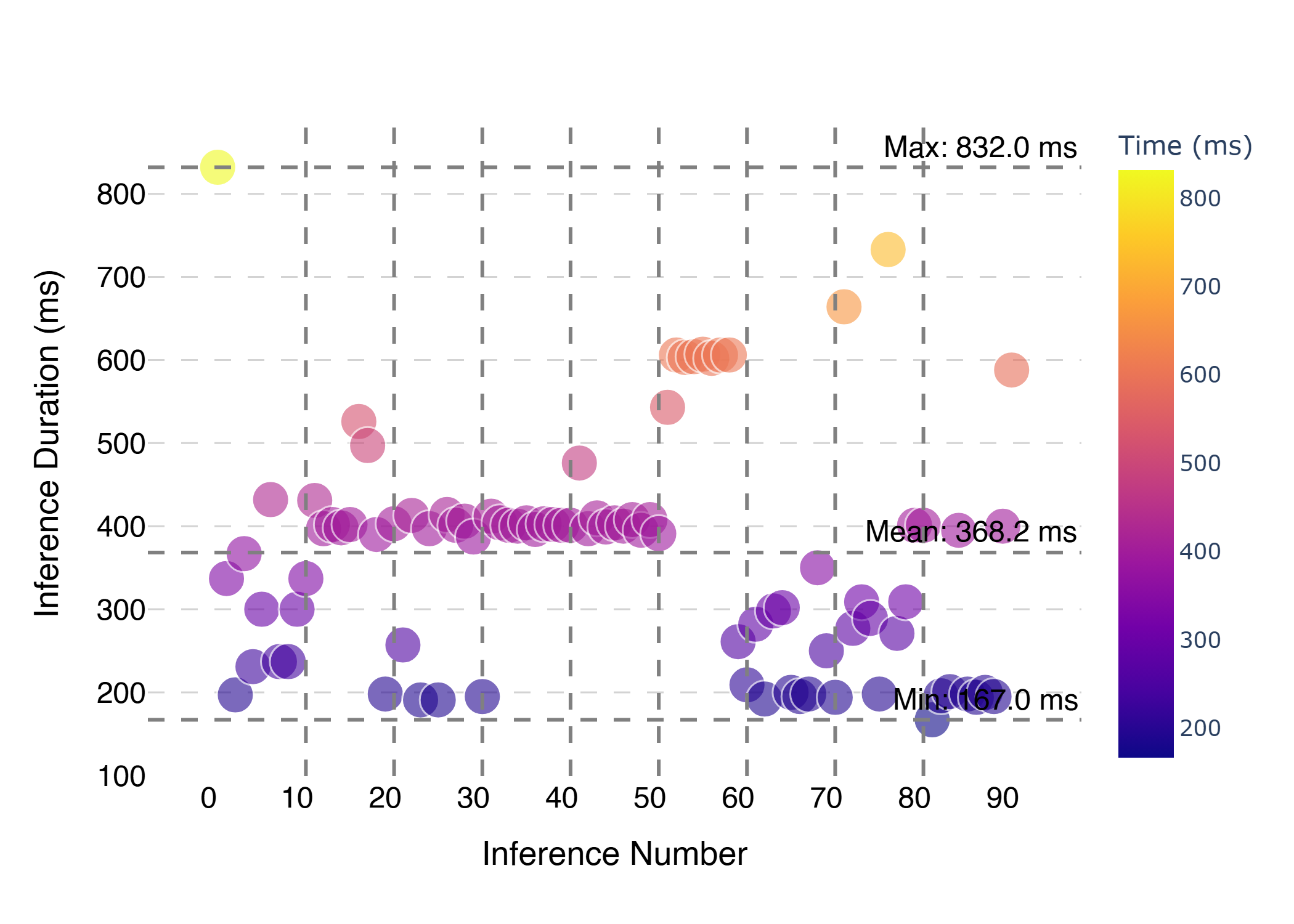}
    \caption{}
    \label{fig:topology-inference-duration}
\end{subfigure}
\caption{Distributed \ac{NN} Inference Durations: (a) using the Publish and Subscribe Strategy and (b) using the Topology Strategy.}
\label{fig:distributted-inference-durations}
\end{figure}

The inference cycle duration was evaluated for both distributed and centralized \ac{NN} implementations.
\Cref{fig:distributted-inference-durations} presents the results for the distributed use case, obtained across multiple trials.
Each trial, composed of 10 samples, is delimited by a vertical dotted line and corresponds to a different sequence in which nodes joined the network, resulting in a distinct topology.
The network structure directly influences inference performance: when devices with higher neuron allocations occupy more central positions in the network, communication paths shorten and inference duration decreases.
In the Publish/Subscribe Strategy, the network structure was determined solely by lower-layer mechanisms, without application-level guidance.
Consequently, even if the node with the highest neuron allocation joined first, it might not become the central hub, as these mechanisms prioritise balancing the network.
In contrast, the Topology Strategy allowed the application to favour parent connections to nodes with higher computational capacity.
This distinction is evident in the results.
In the Publish/Subscribe Strategy (\Cref{fig:pubsub-inference-duration,fig:pubsub-inference-duration}), less efficient connections led to higher inference durations, with times ranging from 272\,ms to 815\,ms and a mean of 526.5\,ms.
The Topology Strategy (\Cref{fig:topology-inference-duration}) achieved a lower mean duration of 368.2\,ms and a minimum of 167\,ms, confirming that application-level control of network topology improves distributed \ac{NN} performance.
Notably, in the Topology Strategy, trials where the Raspberry~Pi joined the network early achieved the best inference durations, as it assumed a central position in the topology.
In both strategies, fluctuations in inference duration within trials align with the high variability observed in the RTT results.

%

\begin{figure}[H]
    \centering
    \includegraphics[width=\linewidth]{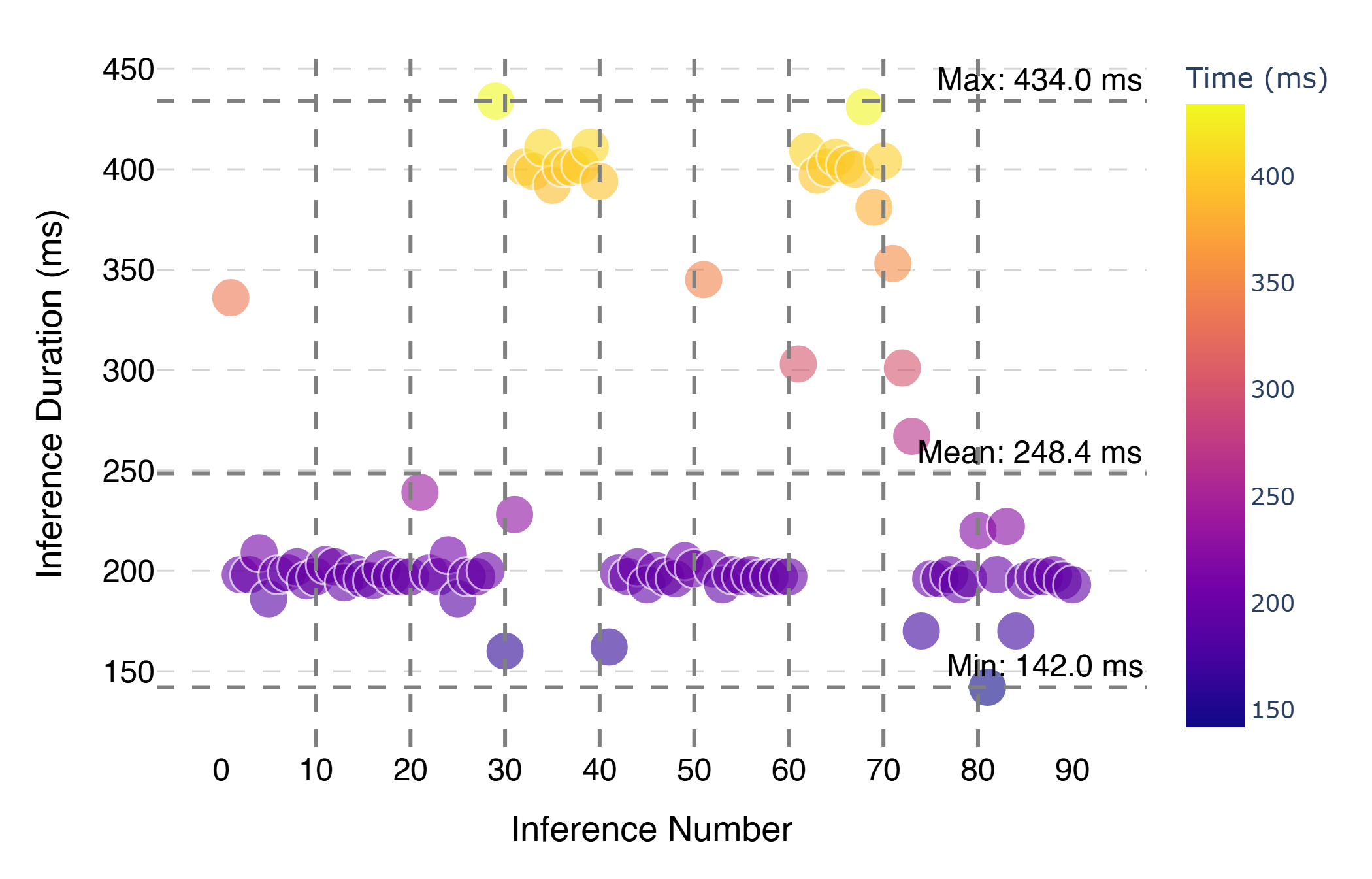}
    \caption{Centralized NN Inference Duration using the Inject Strategy.}
    \label{fig:inference-d-nn-12-inject}
\end{figure}

For the centralized implementation using the Inject Strategy (\Cref{fig:inference-d-nn-12-inject}), all computations are performed locally on the Raspberry~Pi, resulting in the lowest mean inference duration of 248.4\,ms.
Maximum recorded times reached 434\,ms, while the minimum of 142\,ms was close to the best-case performance observed in the distributed scenario using the Topology Strategy.

Network throughput was measured at the root node, as summarized in \Cref{tab:table-throughput-results-app}.
The middleware was configured with a 120\,s update period.
For the distributed \ac{NN} scenario, the Topology Strategy exhibited lower total throughput than the Publish and Subscribe Strategy.
This reduction was most pronounced in application-layer traffic.
Notably, the throughput for Neuron Output messages not destined for the root node was approximately 37\% lower under the Topology Strategy, indicating that its mechanisms optimize application-layer communication by establishing more direct data paths, thereby reducing the number of messages relayed through the root node.

As expected, the centralized scenario with the Inject Strategy exhibited the lowest total throughput, with minimal middleware traffic since only the Raspberry Pi participated in periodic metric updates.

\begin{table}[H]
\centering
\caption{Throughput Results While Having Both the Application and Middleware Layers.}
\scriptsize
\begin{tabular}{lccc}
\toprule
 \textbf{Category} & \textbf{Publish/Subscribe} & \textbf{Topology} & \textbf{Inject} \\
\midrule
Total      & 7.89\,B/s & 6.82\,B/s & 4.22\,B/s \\
Data Total &  5.32\,B/s &  4.18\,B/s & 2.70\,B/s \\
\quad Neuron Output  & 1.62\,B/s & 1.62\,B/s & 0.81\,B/s \\
\quad Neuron Output (forward) & 3.07\,B/s  & 1.94\,B/s & 1.38\,B/s \\
Middleware &  1.35\,B/s &  1.41\,B/s & 0.25\,B/s   \\
Routing    &  0.80\,B/s & 0.80\,B/s & 0.80\,B/s  \\
\bottomrule
\end{tabular}\label{tab:table-throughput-results-app}
\end{table}
\normalsize

During all inference trials, no NACKs were observed across any of the three strategies, indicating that all messages carrying neuron outputs were successfully delivered and all strategies consistently produced correct results.



\section{Conclusions}\label{sec:concl}

This work addressed the challenge of improving routing flexibility in heterogeneous \ac{IoT} networks, aiming to enable applications to achieve their objectives without being constrained by traditional network assumptions.
Experimental validation on a physical testbed highlighted the impact of device heterogeneity on network performance and the influence of the middleware strategies on application behaviour.

The Publish and Subscribe and Inject Strategies provide the advantage that the middleware layer manages all aspects of message forwarding, thereby abstracting and simplifying this complexity for the application layer.
In contrast, the Topology Strategy optimizes network connections, enabling the application layer to operate more efficiently at a different level, but it requires the application to handle raw forwarding details, which can vary and be difficult to manage.
For the distributed use case, the Topology Strategy achieved the best overall performance.
These results suggest that combining the network-level optimization of the Topology Strategy with the message-forwarding simplicity of the Publish/Subscribe and Inject Strategies could provide further improvements and represents a promising direction for future work.

In addition, the distributed \ac{NN} solution developed around a coordinator and multiple worker nodes proved to operate effectively and to remain resilient in the presence of message losses, including missing acknowledgments and neuron outputs, allowing the system to continue functioning amidst such faults.
The experiments also showed that the underlying network topology plays a significant role in the duration of the inference.
Topologies that more closely reflect the structure of the \ac{NN} by aligning the assigned neurons in a way that shortens communication paths consistently achieved better performance.

It is also important to acknowledge the limitations of this work.
The system does not consider the energy constraints inherent to battery-powered nodes, nor does it incorporate task scheduling or power management mechanisms.
Although this is indeed a limitation, it is mitigated in \ac{IoT} applications where nodes are assumed to be connected to the electrical network, and ongoing research is exploring wireless charging techniques for \ac{IoT} devices.
Another limitation concerns the scale and scope of experimental validation:
the framework was tested with a limited number of devices, which does not fully reflect the complexity of large-scale \ac{IoT} deployments.
Additionally, the \ac{NN} case study demonstrates feasibility but is limited to relatively simple architectures, leaving more complex or larger-scale networks unexplored.

The highlighted constraints provide valuable insights into potential directions for future research.
Three principal lines of further development are suggested.
The first concerns improvements to the overall system, particularly in terms of energy efficiency, for example by introducing task scheduling mechanisms and sleep modes.
The second line targets the middleware layer.
The Topology Strategy could be extended to allow modifications not only when devices join but also dynamically during execution.
Another improvement would be to integrate the Topology Strategy with the other two strategies already implemented.
Additional future work concerns the distributed computation of the \ac{NN} across \ac{IoT} devices.
The current system could be further explored through more extensive testing.
In the Topology Strategy, the parent-selection function could be optimized.
Likewise, the algorithm used by the coordinator to distribute the hidden neurons could be improved by adopting a more sophisticated mapping algorithm.

Overall, this work contributes an application-aware routing framework, whose feasibility was demonstrated by distributing \ac{NN} inference across physical devices, showing its capability to support complex edge computing applications.


\bibliographystyle{elsarticle-num}
\bibliography{main}


%

\end{document}